\documentclass[a4paper]{llncs}
\usepackage[latin1]{inputenc}
\usepackage{mathptmx}
\usepackage{latexsym}
\usepackage{amsfonts,amssymb,latexsym}
\usepackage{amsmath}
\usepackage{algorithm}
\usepackage[noend]{algpseudocode}
\usepackage[all]{xy}
\usepackage{verbatim} 
\usepackage{url}
\usepackage[T1]{fontenc}

\usepackage{geometry} 
\usepackage{color} 
\definecolor{Blue}{rgb}{0.1,0.1,0.9}
\definecolor{Red}{rgb}{0.9,0.1,0.1}
\definecolor{Green}{rgb}{0.1,0.9,0.1}

\newcommand{\maria}[1]{{\color{blue}\textbf{{\marginpar{** Maria **}}#1}}}%
\definecolor{copper}{rgb}{0.72, 0.45, 0.2}
\newcommand{\santi}[1]{{\color{copper}\textbf{{\marginpar{** Santi **}}#1}}}%
\newcommand{\angel}[1]{{\color{red}\textbf{{\marginpar{** Angel **}}#1}}}%

\newcommand{\Symbols}{\Sigma}
\newcommand{\VNfold}[1][\caR]{\textit{VN}_{#1}^{\,\fold}}
\newcommand{\Full}[1][\cR]{\textit{Full}_{#1}}
\newcommand{\frontier}[3]{\mathit{Frontier}^{#2}_{{#1}\!\!}(#3)}
\newcommand{\xor}[0]{\oplus}
\newcommand{\narrow}[1]{\leadsto_{#1}}
\newcommand{\narrowG}[2]{\leadsto_{{#1}}^{#2}}
\newcommand{\fold}{\circlearrowleft}
\newcommand{\domain}[1]{\mathit{Dom}(#1)}
\newcommand{\range}[1]{\intrvar{#1}}
\newcommand{\intrvar}[1]{\mathit{Ran}(#1)}

\newcommand{\csuV}[3]{\textit{CSU\/}^{#2}_{#3}({#1})}

\newcommand{\restrict}[2]{#1|_{#2}}
\let\oldAA\AA
\renewcommand{\AA}{\text{\normalfont\oldAA}}

\def\defemb#1#2{\expandafter\def\csname #1\endcsname
	{\relax\ifmmode #2\else\hbox{$#2$}\fi}}

\def\cC{\mathcal{C}}
\def\cD{\mathcal{D}}
\def\cE{\mathcal{E}}

\def\cL{\mathcal{L}}

\def\cN{\mathcal{N}}
\def\cV{\mathcal{V}}

\def\cP{\mathcal{P}}
\def\cR{\mathcal{R}}
\def\caR{\mathcal{R}}
\def\cS{\mathcal{S}}
\def\caS{\mathcal{S}}

\def\cX{\mathcal{X}}

\def \tuple#1{\langle #1 \rangle}

\long\def\comment#1{}

\newcommand{\idsubst}{\textit{id}}
\newcommand{\var}{{\cV}ar}

\newcommand{\dom}{{\cD}om}

\def\defemb#1#2{\expandafter\def\csname #1\endcsname
	{\relax\ifmmode #2\else\hbox{$#2$}\fi}}

\def\ll{[\![}
\def\rr{]\!]}
\def\Den#1{\relax\ifmmode \ll #1\rr \else\hbox{$\ll #1\rr$}\fi}

\let\|=\mid

\def\#{\hat{~}}

\defemb{implies}{\quad\rightarrow\quad}
\defemb{Implies}{\quad\Rightarrow\quad}

\defemb{cA}{{\cal A}}
\defemb{cB}{{\cal B}}
\defemb{cC}{{\cal C}}
\defemb{cD}{{\cal D}}
\defemb{cE}{{\cal E}}
\defemb{cF}{{\cal F}}
\defemb{cG}{{\cal G}}
\defemb{cH}{{\cal H}}
\defemb{cI}{{\cal I}}
\defemb{cJ}{{\cal J}}
\defemb{cL}{{\cal L}}
\defemb{cP}{{\cal P}}
\defemb{cR}{{\cal R}}
\defemb{cS}{{\cal S}}
\defemb{cT}{{\cal T}}
\defemb{caT}{{\cal T}}
\defemb{cU}{{\cal U}}
\defemb{cV}{{\cal V}}
\defemb{cX}{{\cal X}}
\defemb{caX}{{\cal X}}
\defemb{cZ}{{\cal Z}}

\def \tuple#1{\langle #1 \rangle}

\newcommand*{\ul}[1]{\underline{#1}}

\long\def\comment#1{}

\defemb{cR}{{\cal R}}
\defemb{cE}{{\cal E}}
\defemb{cN}{{\cal N}}
\defemb{cV}{{\cal V}}
\defemb{cP}{{\cal P}}
\defemb{cX}{{\cal X}}
\defemb{cU}{{\cal U}}

\newcommand{\Xc}{{\mathcal{X}}}
\newcommand{\Tc}{{\mathcal{T}}}
\newcommand{\Var}{{\mathcal V}ar}

\newcommand{\ol}[1]{\overline{#1}}
\newcommand{\toppos}{\mbox{\footnotesize$\Lambda$}}
\newcommand{\pos}{{\cP}os}
\newcommand{\fpos}{{\cN}{\cV}{\cP}os}

\newcommand{\rewrite}[1]{\rightarrow_{#1}}
\newcommand{\rewrites}[1]{\rightarrow^*_{#1}}

\newcommand{\TermsOn}[5]{{\caT^{#4}_{#1}(#2)}_{#3}^{#5}}
\newcommand{\Terms}{\TermsOn{\Symbols}{\Variables}{}{}{}}
\newcommand{\Variables}{\caX}
\newcommand{\funocc}[1]{\mathit{Pos}_{\Symbols}(#1)}
\newcommand{\subterm}[2]{#1|_{#2}} 
\newcommand{\congr}[1]{=_{\protect #1}}
\newcommand{\replace}[3]{#1[#3]_{#2}}
\newcommand{\norm}[1]{{\downarrow_ {#1}}}
\newcommand{\TermsS}[1]{\TermsOn{\Symbols}{\Variables}{\sort{#1}}{}{}}
\newcommand{\sort}[1]{\ensuremath{\mathsf{#1}}}
\newcommand{\GTermsOn}[2]{\caT^{#2}_{#1}}
\newcommand{\GTermsS}[1]{\GTermsOn{\Symbols,\sort{#1}}{}}

\newcommand\bcmdtab{\noindent\bgroup\tabcolsep=0pt%
  \begin{tabular}{@{}p{10pc}@{}p{20pc}@{}}}
\newcommand\ecmdtab{\end{tabular}\egroup}

\begin{document}
  \title{Partial Evaluation of Order-sorted Equational Programs modulo Axioms 
  \thanks{This work has been
        		partially supported by the EU (FEDER) and the Spanish
        		MINECO under grants TIN 2015-69175-C4-1-R
        		and TIN 2013-45732-C4-1-P,
        		and by Generalitat Valenciana under grant PROMETEOII/2015/013, and by NSF grant CNS-1319109.}}

  \author{
  Mar\'{\i}a Alpuente\inst{1}
  \and
  Angel Cuenca\inst{1}
  \and
  Santiago Escobar\inst{1}
  \and
  Jos\'e Meseguer\inst{2}
}

\institute{
DSIC-ELP, Universitat Polit\`ecnica de Val\`encia, Spain.
\email{\{alpuente,acuenca,sescobar\}@dsic.upv.es}
\and
University of Illinois at Urbana-Champaign, USA.
\email{meseguer@illinois.edu}
}



\maketitle

  \begin{abstract}
    Partial evaluation (PE) is a powerful and general program optimization technique with many successful applications.
However, it has never been investigated in the context of expressive rule-based languages like Maude, CafeOBJ, OBJ, ASF+SDF, and ELAN, which  support: 1) rich type structures with sorts, subsorts and overloading; 2) equational rewriting modulo axioms such as commutativity, associativity--commutativity, and 
    associativity--commutativity--identity.
    In this 
    extended abstract, we 
    illustrate the key concepts by showing how they apply to partial evaluation of expressive rule-based programs written in Maude.
    Our partial evaluation scheme is based on an automatic
    unfolding algorithm that computes term {\em variants} and relies on {\em equational} least general generalization for ensuring global termination.
    We demonstrate the use   of the resulting partial evaluator for program optimization
    on several examples where it
    shows significant speed-ups.
  \end{abstract}



\pagestyle{plain}

\section{Introduction}\label{sec:intro}


Partial evaluation (PE) is a semantics-based program transformation
technique in which a program is specialized to a part of its input that is known statically (at {\em specialization} time)  \cite{DagstuhlPE96,JGS93}. 
Partial evaluation has currently reached a point where theory and refinements 
have matured, substantial systems have been developed, and realistic applications can benefit from partial evaluation in a wide range of   fields that transcend by far 
program optimization. 

Narrowing-driven PE (NPE) 
\cite{
AFV98,AFV98b} 
is a generic algorithm for 
the specialization of  functional programs that are executed by {\em narrowing} \cite{Fay79,Sla74}, 
an extension of rewriting where matching is replaced by unification.
Essentially, narrowing consists of computing an appropriate substitution for a symbolic program call 
in such a way that the program call becomes reducible,
and then reduce it:  both the rewrite rule and the term can be instantiated. 
As in logic programming, narrowing computations can be represented by a  (possibly infinite) finitely branching tree.
Since
narrowing subsumes both rewriting and SLD-resolution, it is complete in the
sense of both functional programming (computation of normal forms) and 
logic programming (computation of answers). 
%
%
%
%
By combining the  functional dimension of narrowing with the power of logic variables and unification, the NPE approach has better opportunities for optimization than the more standard 
partial evaluation of logic programs (also known as {\em partial deduction\/}, PD) and functional programs   \cite{AFV98b}.


To the best of our knowledge, partial evaluation has never been investigated in the context of expressive rule-based languages like Maude, CafeOBJ, OBJ, ASF+SDF, and ELAN, which  support: 1) rich type structures with sorts, subsorts and overloading; 
and 2) equational rewriting modulo axioms such as commutativity, associativity--commutativity, and 
    associativity--commutativity--identity.
    In this 
    extended abstract, 
    we 
    illustrate the key concepts by showing how they apply to partial evaluation of expressive rule-based programs written in Maude.
The key NPE ingredients  of \cite{AFV98} have to be further generalized to corresponding  (order--sorted)  {\em equational} notions ({\em modulo} axioms): e.g.,\ {\em equational unfolding, equational closedness, equational embedding, and equational 
abstraction};
and the associated partial evaluation techniques become more sophisticated and powerful.


Let us motivate the power of our technique by reproducing the classical specialization of a program parser w.r.t.\ a
given grammar into a very specialized parser
 \cite{JGS93}.


\begin{example} \label{ex:parser1}
	Consider the following rewrite theory (written in Maude\footnote{In Maude 2.7, 
	only equations with the attribute \texttt{variant} are used by
	the folding variant narrowing strategy, which is the only narrowing strategy considered in this paper.} syntax) 
	that defines an elementary parser for the language generated by simple, right regular grammars.
	We define a symbol \verb!_|_|_! to represent the parser configurations,
	where the first underscore represents the  (terminal or non-terminal)   symbol     being processed,
	the second underscore represents the  current  string   pending  to be recognised,
	and the third underscore stands for the considered grammar. 
	We provide two non-terminal symbols \texttt{init} and \texttt{S}
	and 
	three terminal symbols $0$, $1$, and the finalizing mark \texttt{eps} (for $\epsilon$, the empty string).
	These are useful choices for this example, but they can be easily extended to more terminal and non-terminal symbols.
	Parsing a   string $st$ according to a given   grammar $\Gamma$ is defined by rewriting the configuration  
	($  {\tt init \ \mid \ st}$ \nolinebreak ${\tt \mid}$ \nolinebreak\  $\Gamma$) 
	using the rules of the grammar (in the opposite direction)
	to incrementally transform   $st$   until  the final configuration 
	{(\tt eps $\mid$ eps $\mid \Gamma$)} is reached. 

{
	\small	
\begin{verbatim}
	fmod Parser is 
	  sorts Symbol NSymbol TSymbol String Production Grammar Parsing . 
	  subsort Production < Grammar . 
	  subsort TSymbol < String .		  
	  subsorts TSymbol NSymbol < Symbol . 
	  ops 0 1 eps : -> TSymbol . ops init S : -> NSymbol . op mt : -> Grammar .
	  op __ : TSymbol String -> String [right id: eps]. 
	  op _->_ : NSymbol TSymbol -> Production . 
	  op _->_._ : NSymbol TSymbol NSymbol -> Production . 
	  op _;_ : Grammar Grammar -> Grammar [assoc comm id: mt] .
	  op _|_|_ : Symbol String Grammar -> Parsing .	
	  var E : TSymbol . vars N M : NSymbol . var  L : String . var G : Grammar .   
	  eq (N | eps | ( N ->  eps ) ; G) =  (eps | eps | ( N ->  eps ) ; G) [variant] .      
	  eq (N | E L | ( N -> E . M ) ; G) = (M | L   | ( N -> E . M ) ; G) [variant] .
	endfm
	\end{verbatim}
}	
	
	
	Note that this Maude equational program theory contains several novel features that are \emph{unknown land} for (narrowing-driven) partial evaluation:
	1) a subsorting relation \texttt{TSymbol} \texttt{NSymbol} $<$  \texttt{Symbol},  
	and
	2) 
	an associative-commutative with identity symbol \verb!_;_! for representing grammars (meaning that they are handled as a multiset of productions), 
	together with the symbol  \verb!__! with right identity for the input 
	 string.
	The general case of the parser is defined by 
	the second 
	equation that, given the configuration {\tt(N $\mid$ E L $\mid \Gamma$)} where  
	({\tt E L}) is the string to be recognized,  
	searches for 
	the grammar production 
	({\tt N -> E . M}) in $\Gamma$ 
	to recognize symbol  {\tt E}, and 
	proceeds to recognize {\tt L} starting from the non-terminal symbol {\tt M}. 
	Note that the combination of subtypes and equational
	(algebraic) axioms allows for a very compact definition.\\
	

	For example, given the following grammar $ \Gamma$  
	generating the 
	language \texttt{$(0)^{*}(1)^{*}$}:   
{\small

\!\!
	\begin{verbatim}
	    init -> eps    init -> 0 . init   init -> 1 . S    S -> eps    S -> 1 . S
	\end{verbatim}
	}
	
	
	\noindent the initial configuration $(\texttt{init} \ \mid \ 0 \ 0 \ 1 \ 1 \ \tt eps \ \mid \ \Gamma)$
	\comment{
		\texttt{ok} \ . \ 0 \ \texttt{->}\ \texttt{ok}\ ; \ \texttt{init} \ . \ 0 \ \texttt{->}\ \texttt{ok} \ ; \ \texttt{ok} \ . \ 1 \ \texttt{->}\ \texttt{ok}\ ; \ \texttt{init} \ . \ 1 \ \texttt{->}\ \texttt{ok})$
	} 
	is deterministically rewritten as\\
	{\small $(\texttt{init} \ \mid \ 0 \ 0 \ 1 \ 1 \ \tt eps \ \mid \Gamma) \rightarrow (\texttt{init} \ \mid  \ 0 \ 1 \ 1 \ \tt eps \ \mid \Gamma) \ \rightarrow \ {(\texttt{init} \ \mid \ 1 \ 1 \ eps \ \mid \Gamma)} \ \rightarrow \linebreak (\texttt{S} \ \mid \ 1 \ \tt eps \ \mid \Gamma) \ \rightarrow (\texttt{S} \ \mid \ \tt eps \ \mid \Gamma) \rightarrow (\texttt{eps} \ \mid \ \tt eps \ \mid \Gamma)$}. 
\\
	

	
	We can specialize our parsing program to the productions of the given grammar $\Gamma$ by partially evaluating the input term 
	$(\texttt{init} \ \mid \ \tt L \mid \Gamma)$, where \texttt{L} is a logical variable of sort \texttt{String}.
	\comment{\ \texttt{ok} \ \mid \ 0 \ \texttt{->}\ \texttt{Exp}\ ; \ \texttt{init} \ \mid \ 0 \ \texttt{->}\ \texttt{ok} \ ; \ \texttt{Exp} \ \mid \ 1 \ \texttt{->}\ \texttt{Exp}\ ; \ \texttt{init} \ \mid \ 1 \ \texttt{->}\ \texttt{ok}\]
	}
%
%
	By applying our partial evaluator,  we aim to obtain  the   specialized parsing equations: 
	
	\comment{
		\begin{verbatim}
		eq init || eps = eps || eps .
		eq init || 0 L = init || L .
		eq init ||1 L = S || L .
		eq S || eps = eps || eps .
		eq S || 1 L = S || L .	
		\end{verbatim}
	}
{
\small	
	
\begin{verbatim}
 eq init || eps   = eps || eps .       eq init || 1 = eps || eps .       
 eq init || 0 L   = init || L .        eq S || eps  = eps || eps .	     
 eq init || 1 1 L = S || L .           eq S || 1 L  = S || L .
\end{verbatim}
}	

	\noindent
	which get rid of the grammar $\Gamma$ (and hence of costly \emph{ACU-matching} operations) while still recognizing string $st$ by rewriting  the simpler configuration ($  {\tt init {\tt }} \ || \ st$)   to the final configuration ($  {\tt eps} \ || \ {\tt eps}$).
	We have run some test on both the original and the specialized programs with an impressive improvement in performance,
	see Section~\ref{sec:experiments}. 
	\end{example}
	
%
	

	\paragraph{\bf Our contribution.} 
	In this extended abstract, we delve into the essential ingredients of a
        partial evaluation framework 
for	order sorted equational theories that is able to cope with  
	subsorts, subsort polymorphism,  convergent rules (equations), and 
	equational axioms. 
	We base our partial evaluator on 
	a suitably extended version of
	the general  NPE procedure of \cite{AFV98}, which is  
	parametric w.r.t.\  the {\em unfolding rule}  used to construct  finite computation trees
	and also w.r.t.\ an {\em  abstraction operator} that is used to guarantee that only finitely many expressions are evaluated. 
	For unfolding we use {\em (folding) variant narrowing} \cite{ESM12}, a novel narrowing strategy 
	for 
	convergent 
	equational theories  that  computes {\em most general variants} modulo algebraic axioms and is efficiently implemented in Maude.
	For the abstraction we rely on the {\em (order-sorted) equational least general generalization} recently investigated in  \cite{AEEM2014}.

	
	\section{Specializing Equational Theories modulo Axioms}\label{sec:rules}
	In this section, we introduce a   
	partial evaluation algorithm for 
	an equational theory decomposed as a triple 
	$(\Symbols,B,\overrightarrow{E_0})$,  
	where 
	$\Sigma$ is the signature, $E_0$ is a set of convergent   (equations that are implicitly oriented as) rewrite rules and $B$ is a set of commonly occurring axioms such as
	associativity, commutativity, and identity. 
	%
	%
%
	Let us start by recalling the key ideas of the NPE approach.
	We assume the reader is acquainted with the basic notions of term rewriting, Rewriting Logic, and Maude (see, e.g,\ \cite{maude-manual}).
	
	\subsection{The NPE Approach}
	Given a set  $R$ of rewrite rules and a set $Q$ of program calls (i.e.\ input terms), 
	the aim of NPE 
	\cite{
	AFV98} 
	is to derive a new set of rules  $R'$ (called a partial evaluation of $R$ w.r.t.\ $Q$, or a partial evaluation of $Q$ in $R$)
	which computes the same  
	answers and irreducible forms (w.r.t.\ narrowing) than $R$ for any   term that $t$ is inductively covered ({\em closed}) by the  calls in $Q$. This means that every  subterm in the leaves of the execution tree for $t$ in  $R$ that can be narrowed (modulo $B$) in $R$
 can also be narrowed (modulo $B$) in  $R'$. 
	Roughly speaking,  $R'$ is obtained by first constructing a {\em finite} (possibly partial) narrowing tree for the input term $t$, and then gathering together the set of 
	{\em resultants\/} $t\theta_{1} \rightarrow t_{1}$,\ldots, $t\theta_{k} \rightarrow t_{k}$ that can be constructed by considering
	the leaves
	of the tree, say $t_{1},\ldots,t_{k}$, 
	and the    computed substitutions $\theta_{1},\ldots,\theta_{k}$ of  the associated  branches of the tree (i.e.,\ a resultant 
	rule is associated to each root-to-leaf derivation of the   narrowing tree). Resultants 
	perform what in fact is an $n$-step computation in $R$, with $n>0$, by means of a single step computation in $R'$.
	The unfolding process is iteratively
	repeated for every 
	narrowable subterm  of $t_{1},\ldots, t_{k}$  that is not covered by the root nodes of the  already  deployed narrowing trees.  This ensures that resultants form a complete description 
	 covering
	 all calls that may occur at run-time in $R'$. 

	Let us illustrate the classical NPE method with the following example that illustrates its ability 
	to perform {\em deforestation\/} \cite{Wad90}, a popular transformation that neither standard partial evaluation nor 
	partial deduction can achieve \cite{AFV98}. Essentially, 
	the aim of deforestation is to eliminate useless intermediate 
	data structures, thus reducing the number of passes over data. 
	
	\begin{example}\label{ex:flip}
		Consider the following Maude program that computes the mirror image of a (non-empty) binary tree, which is built with 
		the free constructor 
		\verb!_{_}_! 
		that stores an element as root above two given (sub-)trees, its left
		and right children.
		Note that the program does not contain  any
		equational attributes either for {\tt \_\{\_\}\_} or for the    operation {\tt flip} defined therein:
		
		
{
	\small
	\begin{verbatim}
  fmod FLIP-TREE is protecting NAT . 
   sort NatTree . subsort Nat < NatTree .  vars R L : NatTree . var N : Nat .
   op _{_}_ : NatTree Nat NatTree -> NatTree .  op flip : NatTree -> NatTree .   
   eq flip(N) = N [variant] .  eq flip(L {N} R) = flip(R) {N} flip(L) [variant] .
  endfm
\end{verbatim}
}
		
	\begin{figure}[t]
		%
		
		\centering
		{\scriptsize
		$\xymatrix@R=4ex@C=2ex{
			& 
			\texttt{flip(flip(T))}\ar_{\{\texttt{T} \mapsto \texttt{N}\}}[dl]
			\ar^{\{\texttt{T} \mapsto \texttt{L\ \{N\}\ R}\}}[dr] 
			\\
			\texttt{flip(N)}\ar@{.>}[d] 
			& 
			& 
			\texttt{flip(flip(R) \{N\} flip(L))}\ar@{.>}[d] 
			\\
			\texttt{N} & & \texttt{flip(flip(L)) {N} flip(flip(R))}
		}$
	   }
		\caption{Folding variant narrowing tree for the goal ${\tt flip(flip(T))}$.}
		\label{fig:flipflip} 
	\end{figure}

		%
		
		
		\noindent  By executing the input term \texttt{flip(flip(T))} this program returns the original tree back, but it first computes
		an intermediate, mirrored tree \texttt{flip(T)} of  ${\tt T}$, which is then flipped again. 
		
		Let us partially evaluate the input term  \texttt{flip(flip(T))} following the NPE approach, 
		hence we compute 
		the   folding variant 
		narrowing
		tree depicted\footnote{We show narrowing steps in solid arrows and rewriting steps in dotted arrows.} in
		Figure~\ref{fig:flipflip} 
		for the term {\tt flip(flip(T)))}. 
		This tree does not contain, altogether,    uncovered calls in its leaves. 
		Thus, we get the following residual  program $\cR'$ after introducing the new symbol \texttt{dflip}:


{
	\small
\begin{verbatim}
  eq dflip(N) = N .           eq dflip(L {N} R) = dflip(L) {N} dflip(R) .
\end{verbatim}
}
		%
		
		\noindent
		which  is completely deforested, since the intermediate tree constructed after the first application of
		{\tt flip}
		is not constructed in the residual program using the
		specialised definition of {\tt dflip}. This is equivalent to the program
		generated by deforestation \cite{Wad90}
                 but with a much better performance, see Section~\ref{sec:experiments}.
		Note that the fact that folding variant narrowing \cite{ESM12} ensures  normalization of terms at each step is
		essential for computing the  calls   \texttt{flip(flip(R))} and \texttt{flip(flip(L))}   
		%
		that appear in the rightmost leaf of the   tree in  Figure \ref{fig:flipflip}, which are 
		closed w.r.t.\ the  root node of the tree. 
	\end{example}

	%

	
	
	When we specialize programs that contain sorts, subsorts, rules,  
	 and equational axioms, things get considerably more involved,
	as discussed in the following section.
	\subsection{Partial evaluation of convergent rules modulo axioms}
	
	Let us motivate the problem  by considering the following variant of the \texttt{flip} function of Example \ref{ex:flip}
	for (binary) graphs  instead of trees.
	
	\begin{example}\label{ex:graph}
		Consider the following Maude program for flipping binary graphs that are  
		represented as  
		multisets of nodes which  may contain
		explicit, left and right, references (pointers)
		to their child nodes in the graph. 
		We use symbol $\sharp$ to denote an empty pointer.\ 
		As expected, the  {\tt BinGraph} (set) constructor \texttt{\_;\_} obeys axioms of associativity, commutativity and identity (ACU). 
		We consider 
		a fixed set of 
		identifiers 
		$\texttt{0}\ldots\texttt{4}$.
		
		\comment{
			\maria{Meter un supersort {\tt?BinGraph} o subir el operador {\tt flip} al kind {\tt [Node]}. Ahora hay dos operadores {\tt ;}: el del tipo m\'as alto es definido y 
				el del tipo m\'as bajo es constructor. Citar TR U. Illinois \cite{Mes2015}: (J. Meseguer, Order-sorted Rewriting and Congruence Closure, IDEALS repository, UIUC) (LPAR submitted).}
		}
{
	\small
	\begin{verbatim}
  fmod GRAPH is sorts BinGraph Node Id Ref . 
    subsort Node < BinGraph . subsort Id < Ref .
    op {___} : Ref Id Ref -> Node . op mt : -> BinGraph . 
    op _;_ : BinGraph BinGraph -> BinGraph [assoc comm id: mt] .
    ops 0 1 2 3 4 : -> Id . --- Fixed identifiers
    op # : -> Ref . --- Void pointer 
    var I : Id . vars R1 R2 : Ref . var BG : BinGraph .
  endfm   
		\end{verbatim}
		}
		
		\noindent
		We are interested in flipping a graph and define a function \texttt{flip} that takes a reference and a binary graph and returns the flipped graph. 
		
		
{
	\small		
\begin{verbatim}
  op flip : BinGraph -> BinGraph .
  eq [E1] : flip(mt) = mt  [variant] .   
  eq [E2] : flip({R1 I R2} ; BG) = {R2 I R1} ; flip(BG) [variant] .
\end{verbatim}
}		
		
\comment{
		We can represent the following graph
		
		$$\xymatrix@R=2ex{
			& 0\ar[dl]\ar[dr] \\
			1 & & 2\ar[dl]\ar[dr] \\
			& 3\ar[rr]& & 4\ar@/_2pc/[uull]
		}$$
		
		\noindent
		as the following term {\tt BG} of sort \textsf{BinGraph}:
		
		\begin{verbatim}
		{ 1 0 2 } ; { # 1 # } ; { 3 2 4 } ; { # 3 4 } ; { # 4 0 }
		\end{verbatim}
		
		\noindent
		By invoking {\tt flip(BG)} we can flip the above graph {\tt BG}, which delivers the following graph: 
		
		$$\xymatrix@R=2ex{
			& & 0\ar[dl]\ar[dr] \\
			& 2\ar[dl]\ar[dr] & &1 \\
			4\ar@/^2pc/[uurr] & &  3\ar[ll]  
		}$$
}

		We can represent the graph shown on the left-hand side of Figure \ref{side-by-side1}   as the following term {\tt BG} of sort \textsf{BinGraph}:
		{\small
		\begin{verbatim}
		  { 1 0 2 } ; { # 1 # } ; { 3 2 4 } ; { # 3 4 } ; { # 4 0 }
		\end{verbatim} 
	    }
		\noindent		
		By invoking {\tt flip(BG)}, 
		the graph shown on   the right-hand side of Figure \ref{side-by-side1}  is computed.

		\begin{figure}[t] 
			\begin{center}
				\begin{tabular}{|c|c|} 
					\hline 
					\begin{minipage}{0.25\columnwidth}%
						$$\xymatrix @R=1ex @C=2ex{
							& 0\ar[dl]\ar[dr] \\
							1 & & 2\ar[dl]\ar[dr] \\
							& 3\ar[rr]& & 4\ar@/_2pc/[uull]
						}$$
					\end{minipage} & %
					\begin{minipage}{0.25\columnwidth}%
						$$\xymatrix @R=1ex @C=2ex{
							& & 0\ar[dl]\ar[dr] \\
							& 2\ar[dl]\ar[dr] & &1 \\
							4\ar@/^2pc/[uurr] & &  3\ar[ll]  
						}$$				
					\end{minipage}\tabularnewline
					\hline 
				\end{tabular}
			\end{center}	
			\caption{Flipping a graph.}
			\label{side-by-side1}
		\end{figure}
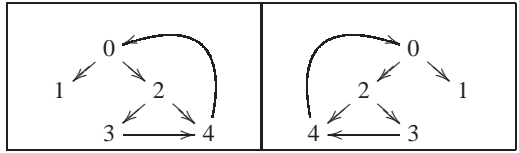 

\end{example}


        In order to 
        specialize the previous program for the call
        \texttt{flip(flip(BG))}, we need several PE ingredients
        that have to be generalized to the corresponding (order--sorted)  {\em equational} notions:
        (i) \emph{equational closedness}, (ii) \emph{equational embedding}, and (iii) \emph{equational generalization}.
       In the following, we  discuss some subtleties about these new notions gradually, through our  graph-flipping  running example.        

%
%
	
	
	\subsection{Equational closedness and the generalized Partial Evaluation scheme}
	
	Roughly speaking, in order to compute a specialization for $t$ in $(\Symbols,B,\overrightarrow{E_0})$,  
	we need to start by   constructing  a finite (possibly partial) 
	$(\overrightarrow{E_0},B)$-narrowing
	tree   for $t$  using the folding variant narrowing strategy~\cite{ESM12}, and then extracting the specialized rules  $t\sigma \Rightarrow r$  (resultants)
	for each narrowing derivation $t \leadsto_{\sigma,\overrightarrow{E_0},B} r$   in the tree.
However, in order to ensure that resultants form a complete description 
covering
all calls that may occur at run-time in the 
final specialized 
theory,
partial evaluation must rely on 
a parametric general notion of \emph{equational $Q$-closedness} (modulo $B$) that is not a mere syntactic subsumption check  (i.e.,\ to be a substitution instance of some term in $Q$  as  in the partial deduction of logic programs), but recurses over the algebraic $B$-structure of the terms.

	\begin{definition}
		[Equational Closedness] \label{closedness}
		Let %
		 $(\Symbols,B,\overrightarrow{E_0})$  
                 be an equational theory decomposition
                 and $Q$ be a finite set of $\Symbols$-terms. 
		Assume the signature $\Sigma$ splits into a set $\cD_{E_0}$ of defined function symbols and a set $\cC_{E_0}$ of constructor symbols (i.e.,\ $\overrightarrow{E_0},B$-irreducible), so that
		$\Sigma = \cD_{E_0} \uplus \cC_{E_0}$. We say that a $\Sigma$-term
		$t$ is closed 
		{\em modulo} $B$ (w.r.t.\ $Q$ and 
		$\Sigma$), or $B$--closed, 
		if $closed_{B}(Q,t)$ 
		holds, where the predicate 
		$closed_{B}$ is defined 
		as follows:
{\small			
		\[ 
		\begin{array}{@{}l}
		closed_{B}(Q,t) \: \Leftrightarrow
		\left\{
		\begin{array}{ll}
		true & \mbox{if } t \mbox{ is a variable} 
		\\
		closed_{B}(Q,t_1)  \wedge  \ldots  \wedge  closed_{B}(Q,t_n) & \mbox{if } t =
		c(\ol{t_n}), 
		~c \in \cC_{E_0}, ~ n\geq 0\\
		\bigwedge_{x\mapsto t' \in
			\theta}
		closed_{B}(Q,t') & \mbox{if } \exists q \in Q 
		\mbox{ such that } q\theta =_{B} t \\
		& \mbox{ for some substitution $\theta$}
		\end{array} \right.
		\end{array}
		\]
}		
		
		\noindent
		A set $T$  of terms is closed   modulo $B$ (w.r.t.\   $Q$ and $\Sigma$) 
		if $closed_{B}(Q,t)$ holds for each $t$ in $T$.
		 A set $R$ of rules is closed   modulo $B$ (w.r.t.\   $Q$ and $\Sigma$) if 
		 the set $\textit{Rhs}(R)$ consisting of the right-hand sides of all the rules in $R$ 
		 is closed   modulo $B$ (w.r.t.\   $Q$ and $\Sigma$).
	\end{definition}

	
	\begin{example}\label{ex:graph-flipflip-BG1}
		In order to partially evaluate the program in Example~\ref{ex:graph} w.r.t.\
		the input term \texttt{flip(flip(BG))},
		we set $Q=\{\texttt{flip(flip(BG))}\}$ and start by constructing the folding variant narrowing tree that is shown\footnote{To ease reading, the arcs of the narrowing tree are decorated with the label of the corresponding equation applied at the narrowing step. 
		}
		in 
		%
		%
		Figure~\ref{fig:graph-flipflip1}.
		
                When we consider the leaves of the tree, we identify two requirements for
                $Q$-closedness, with $B$ being ACU:
		(i) $closed_B(Q,t_1)$ with $t_1=\texttt{mt}$
		and
		(ii) $closed_B(Q,t_2)$ with 
			\linebreak
		$t_2=\texttt{\{R1 I R2\} ; flip(flip(BG'))}$.
		The call $closed_B(Q,t_1)$ holds straightforwardly 
		(i.e.,\ it is reduced to $true$) since the {\tt mt} leaf is a constant and cannot be narrowed. 
		The second one $closed_B(Q,t_2)$ also returns true
		because \texttt{\{R1~I~R2\}} is a flat constructor term and \texttt{flip(flip(BG'))} is a (syntactic) renaming of the root of the tree. 
\end{example}
		
We now show an example  that requires to use $B$-matching in order to ensure equational closedness modulo $B$.
		\comment{
			\begin{figure}[h!]
				{\scriptsize
					\centering
					$\xymatrix@C=5ex{
						& 
						\texttt{flip(R,flip(R,BG))}
						\ar@/_/_>>{\{\texttt{R} \mapsto \sharp\}}[dl]
						\ar_>>{\{\texttt{R} \mapsto \texttt{\^{}I}\}}[d]
						\ar^>>>{\{\texttt{BG} \mapsto \texttt{mt}\}}[dr]
						\ar@/^/|-{\{\texttt{R}\mapsto\texttt{I},\texttt{BG} \mapsto \texttt{\{R1 I R2\} ; BG'}\}}[drr]
						\\
						\texttt{mt}
						&
						\texttt{mt}
						&
						\texttt{flip(R,mt)}
						\ar@{.>}[d]
						&
						\texttt{flip(I,\{R2 I R1\}; flip(R2,flip(R1,BG'))}
						\ar@{.>}[d]
						\\
						&&
						\texttt{mt}
						&
						\texttt{\{R1 I R2\}; flip(R1,flip(R2,flip(R2,flip(R1,BG'))))}
						\ar@{.>}[d]
						\\
						&&&
						\texttt{\{R1 I R2\}; flip(R2,flip(R2,flip(R1,flip(R1,BG'))))}
					}$
				}
				\caption{Folding variant narrowing tree for the goal $\texttt{flip(R,flip(R,BG))}$.}
				\label{fig:graph-flipflip}  
			\end{figure}
			
			\angel{Inicio Graph nuevo}
		}

		\begin{figure}[t]
			\centering
			$\xymatrix@C=3ex@R=6ex{	
				&
				\texttt{flip(flip(BG))}
				\ar|{ 
					\begin{minipage}[t]{1cm}
					\tiny{
						\centering\texttt{$\left[\texttt{E1}\right]$}\\
						$\{\texttt{BG} \mapsto \texttt{mt}\}$
					}\end{minipage}}[dl]
				\ar|{ 
					\begin{minipage}[t]{2cm}
					\tiny{
						\centering\texttt{$\left[\texttt{E2}\right]$}\\
						$\{\texttt{BG} \mapsto \texttt{\{R1 I R2\} ; BG'}\}$
					}\end{minipage}}[dr]
				&
				\\
				\texttt{mt} &&
				\texttt{flip(\{R2 I R1\} ; flip(BG'))}
				\ar@{.>}_{ 
					\begin{minipage}[t]{0.2cm}
					\tiny{
						\centering\texttt{$\left[\texttt{E1}\right]$}\\
						$id$
					}\end{minipage}}[d]
				\\	
				&&\texttt{\{R1 I R2\} ; flip(flip(BG''))}
			}$		
			\caption{Folding variant narrowing tree for the goal \texttt{flip(flip(BG))}.}
			\label{fig:graph-flipflip1}  
		\end{figure}

			\begin{example}\label{ex:graph-fix}
				Let us introduce a new sort \texttt{BinGraph?}
				to encode bogus graphs that may contain spurious nodes in a supersort 
				\texttt{Id?}
				and homomorphically extend the rest of symbols and sorts. 
				For simplicity, we just consider one additional constant symbol \texttt{e} in sort \texttt{Id?}.
%

{
	\small
\begin{verbatim}
  sorts BinGraph? Id? Node? Ref? . subsort BinGraph Node? < BinGraph? . 
  subsort Node < Node? . subsort Id < Id? . subsort Ref Id? < Ref? .
  op e : -> Id? .     op {___} : Ref? Id? Ref? -> Node? . 
  op _;_ : BinGraph? BinGraph? -> BinGraph? [ctor assoc comm id: mt] .
  vars I I1 : Id . var I? : Id? . vars R1 R2 : Ref . vars R1? R2? : Ref? .
  vars BG : BinGraph . var BG? : BinGraph? .     
\end{verbatim}			%
}

\noindent
%
Let us consider a function \texttt{fix} 
that receives an extended graph \texttt{BG?}, an unwanted node \texttt{I?}, and a new content \texttt{I},
and traverses the graph replacing \texttt{I?} by \texttt{I}.
{
	\small
\begin{verbatim}
  op fix : Id Id? BinGraph? -> BinGraph? .
  eq [E3] : fix(I, I?, {R1? I? R2?} ; BG?) = 
            fix(I, I?, {R1? I R2?} ; BG?) [variant] .
  eq [E4] : fix(I, I?, {I? I1 R2?} ; BG?) = 
            fix(I, I?, {I I1 R2?} ; BG?) [variant] .
  eq [E5] : fix(I, I?, {R1? I1 I?} ; BG?) = 
            fix(I, I?, {R1? I1 I} ; BG?) [variant] .
  eq [E6] : fix(I, I?, BG) = BG [variant] .
\end{verbatim}
}

				%
				%
				For example, consider the following term {\tt T} of sort \textsf{BinGraph?}: 
				\begin{verbatim}
				{# 1 e} ; {e 0 #} ; {e e 3} ; {e 3 #} 
				\end{verbatim}
				that represents the graph shown on the left-hand side of Figure \ref{side-by-side2}. By invoking {\tt fix(2, e, T)}, we can fix the graph {\tt T}, by computing the corresponding transformed graph shown   on the right-hand side of Figure \ref{side-by-side2}, where the unwanted node $e$ has been replaced.

				\begin{figure}[t] 
					\begin{center}
						\begin{tabular}{|c|c|} 
							\hline 
							\begin{minipage}{0.25\columnwidth}%
								$$\xymatrix@R=1ex@C=2ex{
									1\ar[dd]& 0\ar[ddl] & \\ & \\
									e\ar@(ul,dl)[] 
									\ar@/^/[dr] &&
									\\
									& 3 \ar@/^/[ul]   
								}$$	
							\end{minipage} & %
							\begin{minipage}{0.25\columnwidth}%
								$$\xymatrix@R=1ex@C=2ex{
									1\ar[dd]& 0\ar[ddl] & \\ & \\
									2\ar@(ul,dl)[] 
									\ar@/^/[dr] &&
									\\
									& 3 \ar@/^/[ul]   
								}$$
							\end{minipage}\tabularnewline
							\hline 
						\end{tabular}
					\end{center} 
								\caption{Fixing a graph.}
			\label{side-by-side2}
				\end{figure}	

				\noindent

\comment{				
				$$\xymatrix@R=2ex{
					1\ar[dd]& 0\ar[ddl] & \\ & \\
					2\ar@(ul,dl)[] 
					\ar@/^/[dr] &&
					\\
					& 3 \ar@/^/[ul]   
				}$$
}

				
				Now assume we want to specialize the above function \texttt{fix} w.r.t. the input term \texttt{fix(2, e, \{R1 I R2\} ; BG?)}, that is, a bogus graph with at least one non-spurious node \texttt{\{R1 I R2\}} (non-spurious because of the sort of variable \texttt{I}). 
Following the proposed methodology, we set $Q=\{\texttt{fix(2, e, \{R1 I R2\} ; BG?)}\}$ and start by constructing the  folding variant narrowing tree shown in 	Figure~\ref{fig:graph-fix3}.   	

			\begin{figure}[t]
							{\scriptsize
							
							$\xymatrix@C=6ex@R=6ex{
								&
								\texttt{fix(2, e, \{R1 I R2\} ; BG?)}
								\ar|{ 
									\begin{minipage}[t]{2.5cm}
									\tiny{
										\centering\texttt{$\left[\texttt{E3}\right]$}\\
										$\{\texttt{BG?} \mapsto \texttt{\{R1?' e R2?'\} ;}$\\
										$\texttt{~~~~~~~~~~~BG?'}\}$					
									}\end{minipage}}[ddl]
								\ar|{ 
									\begin{minipage}[t]{2.1cm}
									\tiny{
										\centering\texttt{$\left[\texttt{E4}\right]$}\\
										$\{\texttt{BG?} \mapsto \texttt{\{e I1 R2?'\} ;}$\\
										$\texttt{~~~~~~~~BG?'}\}$					
									}\end{minipage}}[dd]
								\ar|{ 
									\begin{minipage}[t]{2.1cm}
									\tiny{
										\centering\texttt{$\left[\texttt{E5}\right]$}\\
										$\{\texttt{BG?} \mapsto \texttt{\{R1?' I1 e\} ;}$\\
										$\texttt{~~~~~~~~BG?'}\}$										
									}\end{minipage}}[ddr]
								\ar|{ 
									\begin{minipage}[t]{1.1cm}
									\tiny{
										\centering\texttt{$\left[\texttt{E6}\right]$}\\
										$\{\texttt{BG?} \mapsto \texttt{BG}\}$
									}\end{minipage}}[dr]
								&&&						
								\\			
								&
								&				
								\texttt{\texttt{\{R1 I R2\} ; BG}}	&
								\\
								\begin{minipage}[t]{3,3cm}
								\texttt{fix(2, e, \{R1?' 2 R2?'\} ;}
								$~~~~~~~~~~~~~~~~~~~$\texttt{BG?' ; \{R1 I R2\})}
								\end{minipage}											
								&			
								\begin{minipage}[t]{3,2cm}
								\texttt{fix(2, e, \{2 I1 R2?'\} ;} 
								$~~~~~~~~~~~~~~~$\texttt{BG?' ; \{R1 I R2\})}			
								\end{minipage}						
								&
								\begin{minipage}[t]{3,2cm}
								\texttt{fix(2, e, \{R1?' I1 2\} ;}
								$~~~~~~~~~~~~~~~$\texttt{BG?' ; \{R1 I R2\})}			
								\end{minipage}						
								&&&						
							}$
						}
						\caption{Folding variant narrowing tree for the goal $\texttt{fix(2, e, \{R1 I R2\} ; BG?)}$.}
						\label{fig:graph-fix3}  
		\end{figure}

				
		
				The right leaf \texttt{\{R1 I R2\} ; BG} is a 
				constructor term and cannot be unfolded. The first two branches to the left of the tree are closed 
				modulo $ACU$ with the root 
				of the tree in Figure~\ref{fig:graph-fix3}. 
				For instance, for the left leaf 
				$t=\texttt{fix(2, e, \{R1?' 2 R2?'\} ; BG?' ; \{R1 I R2\})}$, the condition $closed_B(Q,t)$ 
				  is reduced\footnote{Note that this is only true because pattern matching modulo $ACU$ is used for testing closedness.} to true
				because $t$ is an instance (modulo $ACU$) 
				of the root node of the tree, 
				and the  subterm $t'=\texttt{(\{R1?' 2 R2?'\} ; BG?')}$ occurring in the 
				 corresponding ($ACU$-)matcher 
			 is a constructor term.
				The other branches can be proved $ACU$-closed with the tree root in a similar way.
				
\comment{

\angel{antes}
				Now we specialize function \texttt{fix} w.r.t. input term $\texttt{fix(2, e, \{e e R\} ; BG?)}$, that is, a bogus graph with a left loop on the spurious node identified by \texttt{e}. The corresponding folding variant narrowing tree is shown in Figure \ref{fig:graph-fix4}. Unfortunately this tree does not represent all possibles computations for (any $ACU$-instances of) the input term, %
				since the narrowable redexes occurring in the tree leaves are not a recursive instance of the set of already partially evaluated calls.
				We need to introduce a methodology that recurses over the structure of the terms to ensure that all possible calls are covered by the specialization.\\
}

\end{example}

\begin{example}[Example~\ref{ex:graph-fix} continued]\label{ex:graph-fix2}
\label{ex:flipFix}
		Now let us assume that the function \texttt{flip} is extended to (bogus graphs of sort) \texttt{BinGraph?},
		by extending equations \texttt{E1} and \texttt{E2} in the natural way.
		We specialize the whole program containing 
				functions \texttt{flip} and \texttt{fix} w.r.t. input term \texttt{flip(fix(2, e, flip(BG)))}, that is, 
				take a graph \texttt{BG}, flip it, then fix any occurrence of nodes \texttt{e}, and finally flip it again.
				The corresponding folding variant narrowing tree is shown in Figure \ref{fig:graph-fix5}. Unfortunately this tree does not represent all possible computations for (any $ACU$-instances of) the input term, %
				since the narrowable redexes occurring in the tree leaves are not a recursive instance of the only partially evaluated call so far.
That is, the term \texttt{flip(fix(2, e, flip(BG') ; \{R2 I R1\}))} of the rightmost leaf 
is not $ACU$-closed w.r.t. the root node of the tree.				
				As in NPE, we need to introduce a methodology that recurses (modulo $B$) over the structure of the terms to augment 
				the set of specialized calls  in a controlled way, so as to ensure that all possible calls are covered by the specialization.

			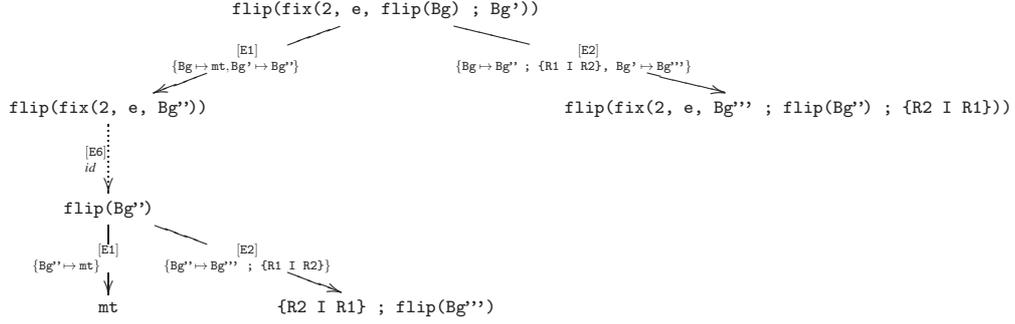
\begin{figure}[t]
				{\scriptsize
					\centering
					$\xymatrix@C=1ex@R=6ex{
						&
						\texttt{flip(fix(2, e, flip(BG)))}
						\ar|{ 
							\begin{minipage}[t]{0.8cm}
							\tiny{
								\centering\texttt{$\left[\texttt{E1}\right]$}\\
								$\{\texttt{BG}\mapsto\texttt{mt}\}$
							}\end{minipage}}[dl]			
						\ar|{ 
							\begin{minipage}[t]{2cm}
							\tiny{
								\centering\texttt{$\left[\texttt{E2}\right]$}\\
								$\{\texttt{BG} \mapsto \texttt{BG' ; \{R1 I R2\}}\}$
							}\end{minipage}}[dr]
						\\
						\texttt{flip(fix(2, e, mt))}
\ar@{.>}_{ 
	\begin{minipage}[t]{0.2cm}
	\tiny{
		\centering\texttt{$\left[\texttt{E6}\right]$}\\
		$id$
	}\end{minipage}}[d]
						&&						
						\texttt{flip(fix(2, e, flip(BG') ; \{R2 I R1\}))}\
						\\
						\texttt{flip(mt)}
\ar@{.>}_{ 
	\begin{minipage}[t]{0.2cm}
	\tiny{
		\centering\texttt{$\left[\texttt{E1}\right]$}\\
		$id$
	}\end{minipage}}[r]
&
						\texttt{mt}
					}$
				}
				\caption{Folding variant narrowing tree for the goal $\texttt{flip(fix(2, e, flip(BG)))}$.}
				\label{fig:graph-fix5}  
			\end{figure}
			
\end{example}

		We are now ready to formulate the backbone of our partial evaluation methodology for 
		equational theories
		that crystallize the ideas of the example above. 
				Following the NPE approach,  we define a generic algorithm (Algorithm \ref{alg:PE-Algorithm}) that is parameterized by:
		\begin{enumerate}
			\item a {\em narrowing relation\/} (with narrowing strategy $\cS$) 
			that constructs search trees,
			\item an {\em unfolding rule\/}, that determines when and how to 
			terminate the construction of the trees, and 
			\item an {\em abstraction operator\/}, that is used to guarantee that the set of terms obtained during partial evaluation 
			(i.e.,\  the set of deployed narrowing trees) is kept finite.
		\end{enumerate}
		
		{\small
		\begin{algorithm}[!h]\caption{Partial Evaluation for Equational Theories}\label{alg:PE-Algorithm}
			\begin{algorithmic}[1]
				\Require
				\Statex An equational theory $\cE = (\Sigma,B,\overrightarrow{E_0})$ 
				and 
				a set of terms $Q$ to be specialized in $\cE$
				\Ensure
				\Statex A set $Q'$ of terms s.t. $\Call{Unfold}{Q',\cE,\cS}$ is closed modulo $B$ w.r.t. $Q'$
				\Function{EqNPE}{$\cR$,$Q$,\cS}
				\State $Q:= Q {\downarrow_{\overrightarrow{E_0},B}}$
				\Repeat
				\State $Q':= Q 
				$
				
				\State $\cL\gets \Call{Unfold}{Q',\cE,\cS}$
				\State $Q\gets \Call{Abstract}{Q'
					,\cL,B}$
				\Until $Q'=_{B} Q$
				\State \Return $Q'$
				\EndFunction
			\end{algorithmic}
		\end{algorithm} }

		\!\!\!
		Informally, the algorithm proceeds as
		follows. Given the input theory $\cE$ and the set of terms $Q$, the first step consists
		in applying the unfolding rule \Call{{\sc Unfold}}{$Q$,$\cE$,\cS} to compute a finite (possibly partial) 
		narrowing tree 
		in $\cE$ for each term $t$ in $Q$, and return the set $\cL$ of the {(normalized)} leaves of the tree. 
		Then,  instead of  proceeding 
		directly with the partial evaluation 
		of the terms  in $\cL$, 
		an abstraction operator $\Call{\sc Abstract}{Q,\cL,B}$  is
		applied that properly combines each uncovered term in $\cL$
		with  the  (already partially evaluated) terms of $Q$, so that the infinite growing of $Q$ is avoided. The abstraction phase yields a new set
		of terms which may need further specialization and, thus, the process is iteratively
		repeated while new terms are introduced.
		
		The PE algorithm does not explicitly compute a partially evaluated theory
		 $\cE'=(\Sigma,B,E')$. It does so implicitly, by computing the set of partially evaluated terms $Q'$ (that unambiguously determine 
			$E'$ 
			as the set of resultants  $t\sigma \Rightarrow r$   
			associated
			to the root-to-leaf derivations  $t \leadsto_{\sigma,\overrightarrow{E_0},B} r$  
			 in the tree, with $t$ in $Q'$), such that the   closedness condition for $E'$ modulo $B$ w.r.t. $Q'$ is satisfied.
	\begin{example}[Example~\ref{ex:graph-fix} continued]\label{ex:graph-fix2}
\label{ex:flipFix}
		Now let us assume that the function \texttt{flip} is extended to (bogus graphs of sort) \texttt{BinGraph?},
		updating  equations \texttt{E1} and \texttt{E2} in the natural way.
		We specialize the whole program containing 
				functions \texttt{flip} and \texttt{fix} w.r.t. input term \texttt{flip(fix(2, e, flip(BG)))}, that is, 
				take a graph \texttt{BG}, flip it, then fix any occurrence of nodes \texttt{e}, and finally flip it again.
				The corresponding folding variant narrowing tree is shown in Figure \ref{fig:graph-fix5}. Unfortunately this tree does not represent all possible computations for (any $ACU$-instances of) the input term, %
				since the narrowable redexes occurring in the tree leaves are not a recursive instance of the only partially evaluated call so far.
That is, the term \texttt{flip(fix(2, e, flip(BG') ; \{R2 I R1\}))} of the rightmost leaf 
is not $ACU$-closed w.r.t. the root node of the tree.				
				As in NPE, we need to introduce a methodology that recurses (modulo $B$) over the structure of the terms to augment in a controlled way the set of specialized calls, so as  to ensure that all possible calls are covered by the specialization.
							
\end{example}


\subsection{Equational homeomorphic embedding}

		Partial evaluation  involves two 
		classical termination problems: the so-called {\em local\/} termination 
		problem (the termination of unfolding, or how to control and keep the 
		expansion of the narrowing trees 
		finite, which is managed by an unfolding rule), and the {\em global\/} termination (which concerns termination of recursive unfolding, or how to stop recursively constructing more and more narrowing 
		trees).
		
		For local termination, 
		we need to define \emph{equational homeomorphic embedding} 
		by extending the standard notion of
		homeomorphic embedding with order-sorted information and reasoning modulo axioms.
		Embedding is a  structural
		preorder under which a term $t$ is greater than, i.e., it embeds, another
		term $t'$, written as $t \triangleright t'$, if $t'$ can be obtained from $t$ by deleting some
		parts. 

			Embedding relations are very popular to ensure termination of {\em symbolic} transformations because, provided the signature is finite, for every infinite sequence of terms 
		$t_1, t_2, \ldots,$ there exist  $i<j$ such that $t_i \trianglelefteq t_j$.
		Therefore, when iteratively
		computing a sequence $t_1, t_2, \ldots,t_n$,  finiteness of the sequence can be
		guaranteed by using the embedding as a whistle \cite{Leu98}: whenever a new expression $t_{n+1}$ is to
		be added to the sequence, we first check whether $t_{n+1}$ embeds any
		of the expressions already in the sequence. 
		If that is the case, we say that $\trianglelefteq$
		whistles, i.e., it has detected (potential) non-termination and the computation
		has to be stopped. Otherwise, $t_{n+1}$ can be safely added to the sequence and the
		computation 
		proceeds. For instance, if we work modulo commutativity (C), we must stop a sequence where the term 
		$u=${\tt s(s(X+Y)$\ast$(s(X)+0))} occurs after $v=${\tt s(X)$\ast$s(X+Y)}, since $v$ embeds $u$ modulo commutativity of $\ast$.


		\begin{definition}[(order-sorted) equational homeomorphic embedding]\label{def:embedding}
			Let 
			$(\Sigma,B,\overrightarrow{E_0})$ be an equational theory decomposition. Consider the TRS  {Emb}($\Sigma$) that consists of all rewrite rules
%
			$
			f(X_1 : A_1,\ldots,X_n : A_n) \rightarrow X_i : A_i 
			$
			with $f : A_1, \ldots, A_n \rightarrow A$  in $\Sigma$ and $i\in\{1,\ldots,n\}$.
			For terms $u$ and $v$ we write $u ~\triangleright_{B} ~v$ 
			if $u\rightarrow^{+}_ {
				{Emb}(\Sigma)
				/B} v'$ and $v' $ is equal to $v$ up to $B$-renaming
			(i.e.\, $v \;{\stackrel{ren}{=}}_B v'$ iff there is a renaming substitution $\sigma$
			such that $v =_{B} v'\sigma$). The relation $\trianglelefteq_{B}$ is 
			called $B$--embedding (or embedding modulo $B$).
		\end{definition}
		
		By using this notion, we stop a branch $t \leadsto t'$ of a folding variant narrowing tree, if any
		narrowing redex of the leaf $t'$
		is embedded (modulo $B$) by the narrowing redex of a preceding term $u$ in the branch,
		i.e., $u|_p \trianglelefteq_{B}  t'|_q$. 
		
		\begin{example}[Example~\ref{ex:graph-fix2} continued] \label{ex:quasilast}
			Consider again the (partial) folding variant narrowing tree 
			of Figure~\ref{fig:graph-fix5}.
                      The narrowing redex 
			$t=\texttt{flip(fix(2, e, flip(BG') ; \{R2 I R1\})}$ in the right branch of the tree 
			embeds modulo $ACU$ 
			the 
			tree root \linebreak $u=\texttt{flip(fix(2, e, flip(BG)))}$. 
			Since the whistle 
			$u \trianglelefteq_{B}  t$
			blows, 
			the unfolding 
			of this branch 
			is stopped. 
			
		\end{example}

\subsection{Equational abstraction via equational least general generalization}		

		For global termination, PE evaluation relies on an abstraction operation
		to ensure that the iterative construction 
		of a sequence of partial narrowing trees 	 	
		terminates while still guaranteeing that the desired amount of specialization 
		is retained and that the equational closedness condition is reached.
		In order to avoid   constructing  infinite sets, instead of just taking the union of the set $\cL$ of non-closed terms in the leaves of the tree and  the set $Q$ of specialized calls, the sets $Q$ and $\cL$
		are \emph{generalized}. 
		Hence, the abstraction operation returns a safe approximation $A$ of $Q\cup \cL$ so that each
		expression in the set $Q \cup \cL$ is closed w.r.t. $A$. 
		Let us show how we can define a suitable
		 abstraction operator by
		using the notion of  {\em equational least general generalization ($lgg_B$)} \cite{AEEM2014}. 
		Unlike the syntactical, untyped  case, there is in general no  unique
		$lgg_B$ in the framework of \cite{AEEM2014}, due to both the order-sortedness and to the equational axioms.
		Instead, there is a finite, minimal and complete set of ${lgg_B}$'s for any two terms, so that any other
		generalizer has at least one of them as a $B$-instance.

	More precisely, given
		the current set of already specialized calls $Q$, in order to add a set $T$ of new terms,
		the function $\Call{Abstract$^{\fold}$}{Q,T,B}$ of Algorithm \ref{alg:PE-Algorithm} is instantiated  with the following function, which relies on the 
		notion of \emph{best matching terms} (BMT), a proper  generalization of   \cite{AAFJV98}
		that is aimed at avoiding loss of
		specialization due to generalization. Roughly speaking, to determine  the best matching terms for $t$ in a set of terms $U$ w.r.t.,\ $B$, $BMT_B(U,t)$,
		for each $u_i$ in $U$, we compute the set $W_i$ of   $lgg_B$'s of $t$ and $u_i$, and    select the subset $M$ of   minimal upper bounds of the union $\bigcup_{i} W_i$. Then, 
		the term $u_j$ belongs to $BMT_{B}(Q,t)$ if at least one lgg element in the corresponding $W_j$  belongs to $M$.
		
%
			\begin{example} \label{ex:bmt}
		 
				Let $t \equiv g(1) \xor 1 \xor g(Y)$, $U \equiv \{   1 \xor g(X),  X \xor g(1),  X \xor Y \}$, and consider $B$ to consist of the associative-commutative (AC) axioms for $\xor$. 
				To compute the best matching terms for $t$ in $U$, we first compute the sets of $lgg_B$'s of $t$ with each of the terms in $U$:
			{\small
				\[ \begin{array}{lll}
				W_1 =&lgg_{AC}(\{ g(1) \xor 1 \xor g(Y), 1\xor g(X)\})= \{\tuple{\{Z \xor 1, \{Z/g(1) \xor g(Y)\}, \{Z/g(X)\}},\\
				&\tuple{Z \xor g(W)\},  \{Z/1 \xor g(1), W/Y\},  \{Z/1, W/X\}}\\
				W_2 =&lgg_{AC}(\{ g(1) \xor 1 \xor g(Y), X \xor g(1)\}) =\{ \tuple{\{Z \xor g(1)\},\{Z/g(1) \xor g(Y)\},\{Z/X\} }\}\\
				W_3 =&lgg_{AC}(\{ g(1) \xor 1 \xor g(Y), X \xor Y)\}) = \tuple{\{Z \xor W\}, \{Z/1,W/g(1)  \xor g(Y)\}, \{Z/X,W/Y\}}\\
				\end{array} \]
			}
				Now, the set $M$ of
				minimal upper bounds of the 
				set $W_1\cup W_2\cup W_3$  is $M=\{Z \xor 1, Z \xor g(1)\}$  and 
				thus we have: $BMT_{AC}(S,t) = \{1 \xor g(X),  X\xor g(1) \}$.
			 
		\end{example}


\begin{definition}[equational abstraction operator] \label{abstract}
				Let  $Q,T$  be two sets of terms. 
				We define 
				$abstract^{\fold}(Q,T,B)   \: = abs_{B}^{\fold}(Q,T)$, where: 
	{\small				
				\[ \left\{
				\begin{array}{ll}
				abs_{B}^{\fold}(\ldots 
				abs_{B}^{\fold}(Q,t_{1}),\ldots,t_{n})
				& \mbox{if } T \equiv \{t_1,\ldots,t_n\}, n > 0 \\
				Q & \mbox{if } T \equiv \mbox{\O} ~\mbox{ or }~ T \equiv \{X\}, with~X \in \cX \\
				abs_{B}^{\fold}(Q,\{t_1,\ldots,t_n\})
				& \mbox{if } T\equiv \{t\}, with~t \equiv c(t_1,\ldots,t_n),~ c \in \cC_{E_0} \\ 
				generalize_{B}(Q,Q',t)
				& \mbox{if } T\equiv \{t\}, with~t \equiv f(t_1,\ldots,t_n),~ f \in \cD_{E_0} \\ 
				\end{array}
				\right. 
				\]
			}
				where $Q' = \{ t' \in Q \mid   root(t)=root(t') \mbox{ and } t' \trianglelefteq_B t\}$,
				and the function {\em generalize} 
				is:
		{\small		
				\[ \begin{array}{ll}
				generalize_{B}(Q,\mbox{\O},t) & = \: 
				Q \cup \{t\} \\
				generalize_{B}(Q,Q',t) & = \: 
				Q  \mbox{ if } t \mbox{ is } Q-closed \\
				generalize_{B}(Q,Q',t) & = \: 
				abs_{B}^{\fold}(Q\setminus BMT_{B}(Q',t), 
				Q''\downarrow_{\overrightarrow{E_0},B}) 
				\end{array} \]
		}	
				\comment{
					\angel{anterior}
					where $Q'' = \{\{w\}\cup\{x\theta \mid \theta \in\{\theta_1,\theta_2\}, x  \in \dom(\theta)\} \} \mid  q \in BMT_{B}(Q',t), \tuple{w,\{\theta_{1},\theta_{2}\}} \in 
					lgg_B(\{ q,t\})\}$.\\ \\
					\angel{nuevo}
				}
				where $Q'' = \{l \mid  q \in BMT_{B}(Q',t), \tuple{w,\{\theta_{1},\theta_{2}\}} \in 
				lgg_B(\{q,t\}), x  \in \dom(\theta_1) \cup \dom(\theta_2),~ l\in \{w, x\theta_{1},  x\theta_{2}\}\}$.
			\end{definition}

		\begin{figure}[t]
			{\scriptsize
				\centering
				$\xymatrix@C=1ex@R=8ex{
					&
					\texttt{flip(fix(2, e, flip(Bg) ; Bg'))}
					\ar|{ 
						\begin{minipage}[t]{2cm}
						\tiny{
							\centering\texttt{$\left[\texttt{E1}\right]$}\\
							$\{\texttt{Bg}\mapsto\texttt{mt}, 								\texttt{Bg'}\mapsto\texttt{Bg''}\}$
						}\end{minipage}}[dl]			
					\ar|{ 
						\begin{minipage}[t]{3.5cm}
						\tiny{
							\centering\texttt{$\left[\texttt{E2}\right]$}\\
							$\{\texttt{Bg} \mapsto \texttt{Bg'' ; \{R1 I R2\}, }
							\texttt{Bg'} \mapsto \texttt{Bg'''}\}$							
						}\end{minipage}}[dr]
					\\
					\texttt{flip(fix(2, e, Bg''))}
				\ar@{.>}_{ 
					\begin{minipage}[t]{0.2cm}
					\tiny{
						\centering\texttt{$\left[\texttt{E6}\right]$}\\
						$id$
					}\end{minipage}}[d]
					&&						
					\texttt{flip(fix(2, e, Bg''' ; flip(Bg'') ; \{R2 I R1\}))}\	\\
					\texttt{flip(Bg'')}		
					\ar|{ 
						\begin{minipage}[t]{2cm}
						\tiny{
							\centering\texttt{$\left[\texttt{E1}\right]$}\\
							$\{\texttt{Bg''}\mapsto\texttt{mt}\}$
						}\end{minipage}}[d]			
					\ar|{ 
						\begin{minipage}[t]{2.2cm}
						\tiny{
							\centering\texttt{$\left[\texttt{E2}\right]$}\\
							$\{\texttt{Bg''} \mapsto \texttt{Bg''' ; \{R1 I R2\}}\}$							
						}\end{minipage}}[dr]	\\
										\texttt{mt}		&					\texttt{\{R2 I R1\} ;  flip(Bg''')}									
				}$
			}
			\caption{Folding variant narrowing tree for the goal $\texttt{flip(fix(2, e, flip(Bg) ; Bg'))}$.}
			\label{fig:graph-fix6}  
		\end{figure}

		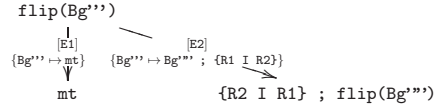
\begin{figure}[t]
			\centering
			{\scriptsize
				$\xymatrix@C=10ex@R=6ex{
					\texttt{flip(Bg''')}		
					\ar|{ 
						\begin{minipage}[t]{1.5cm}
						\tiny{
							\centering\texttt{$\left[\texttt{E1}\right]$}\\
							$\{\texttt{Bg'''}\mapsto\texttt{mt}\}$
						}\end{minipage}}[d]			
					\ar|{ 
						\begin{minipage}[t]{2.3cm}
						\tiny{
							\centering\texttt{$\left[\texttt{E2}\right]$}\\
							$\{\texttt{Bg'''} \mapsto \texttt{Bg'''' ; \{R1 I R2\}}\}$							
						}\end{minipage}}[dr]	\\
					\texttt{mt}		&					\texttt{\{R2 I R1\} ;  flip(Bg'''')}									
				}$
			}
			\caption{Folding variant narrowing tree for the goal $\texttt{flip(Bg''')}$.}
			\label{fig:graph-fix7}  
		\end{figure}

		\begin{example}[Example~\ref{ex:quasilast} continued] \label{ex:quasilast2}
			Consider again the (partial) folding variant narrowing tree 
			of Figure~\ref{fig:graph-fix5}
			with the leaf
						$t=\texttt{flip(fix(2, e, flip(BG') ; \{R2 I R1\}))}$ in the right branch of the tree 
			and the 
			tree root $u=\texttt{flip(fix(2, e, flip(BG)))}$. 			
		We apply the abstraction operator 
		with $Q=\{u\}$ and $T=\{t\}$. 
		Since $t$ is operation-rooted, we 
		call $generalize_{B}(Q,Q',t)$ with $Q'= Q$, which in turn calls 
		$abs_{ACU}^{\fold}(Q \setminus BMT_{ACU}(Q',t), Q'')$, with $BMT_{ACU}(Q',t) = Q$ and $Q''=\{w, v\}$, where $w = \texttt{flip(fix(2, e, flip(Bg) ; Bg'))}$ is the only $ACU$ least general generalization of 
		$u$ and $t$  and $v=\texttt{\{R2' I' R1'\}}$. Then the call returns the set $\{w\}$.
		However, this means that the previous folding narrowing tree of Figure~\ref{fig:graph-fix5}
		is now discarded, since the previous set of input terms $Q=\{u\}$
		is now replaced by $Q'=\{w\}$.

		We start from scratch and the tree resulting for the new call $w$ is showed in Figure~\ref{fig:graph-fix6}. The right leaf embeds the root of the tree and is $B$-closed w.r.t. it. The left leaf \texttt{mt} is a constructor term. For the middle leaf $t'' = \texttt{\{R2 I R1\} ;  \texttt{flip(Bg''')}}$ the whistle $\texttt{flip(Bg'')} \trianglelefteq_{ACU}  t''$ blows and we stop the derivation.
		However, it is not $B$-closed w.r.t. $w$ 
		and we have to add it to the set $Q'$, obtaining the new set of input terms $Q''=\{w,\texttt{flip(Bg''')}\}$.
The specialization of the call \texttt{flip(Bg''')} amounts constructing the narrowing tree of Figure \ref{fig:graph-fix7}, which is trivially $ACU$-closed w.r.t. its root. 

\end{example}

\begin{example}[Example~\ref{ex:quasilast2} continued]\label{ex:quasilast3}
Since the two trees in Figures \ref{fig:graph-fix6} and \ref{fig:graph-fix7} do
			represent all possible computations for (any $ACU$-instance of) 
			$u = \texttt{flip(fix(2, e, flip(BG)))}$,  the partial evaluation process ends.
			Actually $u$ is an instance of the root of the tree in Figure \ref{fig:graph-fix6} with $\{\texttt{Bg'} \mapsto \texttt{mt}\}$ because of the identity axiom. The computed specialization is the set 
			$Q'''$. 
			Now 
			we can extract the set of resultants  $t\sigma \Rightarrow r$   
			associated
			to the root-to-leaf derivations  $t \leadsto_{\sigma,\overrightarrow{E_0},B} r$   in the two trees,
			which yields:
		
		
		{
			\small
		\begin{verbatim}
			  eq flip(fix(2, e, flip(mt))) = mt .
			  eq flip(fix(2, e, flip({R1 I R2} ; BG'))) = 
			     flip(fix(2, e, flip(BG') ; {R2 I R1})) .
			  eq flip(fix(2, e, flip(mt) ; mt)) = mt .
			  eq flip(fix(2, e, flip(mt) ; Bg ; {R1 I R2})) = {R2 I R1} ; flip(Bg) .
			  eq flip(fix(2, e, flip({R1 I R2} ; Bg) ; Bg')) = 
			     flip(fix(2, e, flip(Bg) ; {R2 I R1} ; Bg')) .			
			  eq flip(mt) = mt .
			  eq flip(Bg ; {R1 I R2}) = {R2 I R1} ; flip(Bg) .
		\end{verbatim}
		}
		
		\comment{
			\begin{verbatim}
			eq flip(#,flip(#,BG);X) = null .
			eq flip(^I,flip(^I,BG);X) = null .
			eq flip(I,flip(I,null);X) = flip(I,X) .
			eq flip(I,flip(I,{R1 I R2};BG');X) 
			= {R1 I R2} ; flip(R1,flip(R1,BG');flip(R2,BG');X) 
			; flip(R2,flip(R1,BG');flip(R2,BG');X) .
			eq flip(I,flip(I,BG);{R1 I R2};X') 
			= {R2 I R1} ; flip(I,flip(R1,flip(I,BG);X'); 
			flip(R2,flip(I,BG);X')) .
			
			eq flip(#, BG) = null .
			eq flip(^I, BG) = null .
			eq flip(I, null) = null .
			eq flip(I, {R1 I R2} ; BG) = {R2 I R1} ; flip(R1,BG) ; flip(R2,BG) .
			\end{verbatim}
		} 

\end{example}

The reader may have realized that
the specialization call \texttt{flip(fix(2,e,flip(BG)))} should really return the same term \texttt{BG},
since the variable \texttt{BG} is of sort \texttt{BinGraph} instead of \texttt{BinGraph?},
i.e., $\texttt{flip(fix(2,e,flip(BG)))} = \texttt{BG}$.
The resultants above traverse the given graph and return the same graph. Though the code may seem inefficient,
  we have considered this example to illustrate the different stages of partial evaluation.
The following example shows how a better specialization program can be obtained.
		
	\begin{figure}[t]
		\centering
		{\scriptsize
		$\xymatrix@C=8ex@R=6ex{	
			&
			\texttt{flip(fix(2, e, flip(BG)))}
			\ar@{.>}_{ 
				\begin{minipage}[t]{0.2cm}
				\tiny{
					\centering\texttt{$\left[\texttt{E6}\right]$}\\
					$id$
				}\end{minipage}}[d]	\\
			&
			\texttt{flip(flip(BG))}
			\ar|{ 
				\begin{minipage}[t]{1cm}
				\tiny{
					\centering\texttt{$\left[\texttt{E1}\right]$}\\
					$\{\texttt{BG} \mapsto \texttt{mt}\}$
				}\end{minipage}}[dl]
			\ar|{ 
				\begin{minipage}[t]{2cm}
				\tiny{
					\centering\texttt{$\left[\texttt{E2}\right]$}\\
					$\{\texttt{BG} \mapsto \texttt{\{R1 I R2\} ; BG'}\}$
				}\end{minipage}}[dr]
			&
			\\
			\texttt{mt} &&
			\texttt{flip(\{R2 I R1\} ; flip(BG'))}
			\ar@{.>}_{ 
				\begin{minipage}[t]{0.2cm}
				\tiny{
					\centering\texttt{$\left[\texttt{E1}\right]$}\\
					$id$
				}\end{minipage}}[d]
			\\	
			&&\texttt{\{R1 I R2\} ; flip(flip(BG'))}
		}$
	}		
		\caption{Folding variant narrowing tree for the goal \texttt{flip(fix(2, e, flip(BG)))}.}
		\label{fig:graph-fix8}  
	\end{figure}
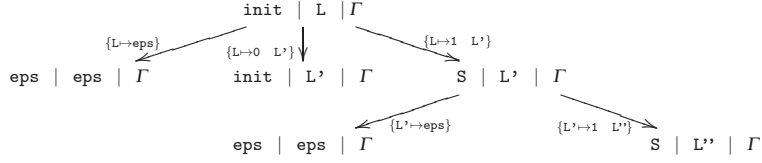

\begin{example}\label{ex:finalEx}
Let us now consider a variant of function \texttt{fix} where its sort is declared as:\\
\verb|  op fix : Id Id? BinGraph? -> BinGraph .|\\
instead of \\ 
\verb|  op fix : Id Id? BinGraph? -> BinGraph? .|\\
Then, if we now specialize the call $t=\texttt{flip(fix(2,e,flip(BG)))}$ in the resulting mutated program, the narrowing tree for $t$ is shown in Figure~\ref{fig:graph-fix8}.
The narrowing tree is $B$-closed w.r.t. the set of calls $\{\texttt{flip(fix(2, e, flip(BG))), flip(flip(BG'))}\}$ (normalized) root of the tree and leads to the following, optimal specialized program:

{
	\small
\begin{verbatim}
eq flip(fix(2,e,flip(mt))) = mt .   
eq flip(fix(2,e,flip({R1 I R2} ; BG))) = {R1 I R2} ; flip(flip(BG)) .   
eq flip(flip(mt)) = mt .      
eq flip(flip({R1 I R2} ; BG)) = {R1 I R2} ; flip(flip(BG)) .
\end{verbatim}
}
\end{example}

		\subsection{Post-processing renaming}
		
The resulting partial evaluations 
		might be further optimized by eliminating	redundant function symbols and unnecessary repetition of variables. 
Essentially, we introduce a new function symbol
		for each specialized term and then replace 
		each call in the specialized program by a call to the corresponding
		renamed function.

		\begin{example}[Example \ref{ex:finalEx} continued] Consider the following independent renaming  for the specialized calls:
		$	\{\texttt{flip(flip(BG))} \mapsto	\texttt{dflip(BG)}, $
		       $\texttt{flip(fix(2,e,flip(BG)))} \mapsto\linebreak	\texttt{dflip-fix(BG)}\}.$
%
				 The post-processing renaming 
				 derives the renamed program				

{
	\small
			\begin{verbatim}
			 eq dflip-fix(mt) = mt .  eq dflip-fix({R1 I R2} ; BG) = {R1 I R2} ; dflip(BG) .				 			
			 eq dflip(mt) = mt . 	    eq dflip({R1 I R2} ; BG') = {R1 I R2} ; dflip(BG') .		
			\end{verbatim}
}			

		\end{example}		
		


		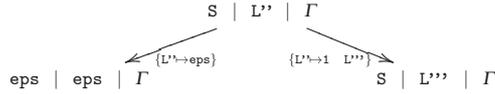
\begin{figure}[t!]
			{\scriptsize
				\centering
				$\xymatrix@C=8ex@R=4ex{
					&&
					\texttt{$\texttt{init} \ \mid \ \texttt{L} \ \mid \Gamma$}
					\ar_>>{\{\texttt{L} \mapsto \texttt{eps}\}}[dl]
					\ar_>>{\{\texttt{L} \mapsto \texttt{0 \ L'}\}}[d]
					\ar^>>>{\{\texttt{L} \mapsto \texttt{1 \ L'}\}}[dr]
					\\
					&
					\texttt{$\texttt{eps} \ \mid \ \texttt{eps} \ \mid \ \Gamma$}
					&
					\texttt{$\texttt{init} \ \mid \ \texttt{L'} \ \mid \ \Gamma$}
					&
					\texttt{$\texttt{S} \ \mid \ \texttt{L'} \ \mid \ \Gamma$}
					\ar^{\{\texttt{L'} \mapsto \texttt{eps}\}}[dl]
					\ar_>>>>>>>{\{\texttt{L'} \mapsto \texttt{1 \ L''}\}}[dr]			
					\\
					&&
					\texttt{$\texttt{eps} \ \mid \ \texttt{eps} \ \mid \ \Gamma$}
					&&
					\texttt{$\texttt{S} \ \mid \ \texttt{L''} \ \mid \ \Gamma$}
				}$
			}
			\caption{Folding variant narrowing tree for the goal $\texttt{init} \ \mid \ \texttt{L} \ \mid \Gamma$.}
			\label{fig:graph-paser}  
			\vspace{-1ex}
		\end{figure}
		
		\begin{figure}[t!]
			\centering
			{\scriptsize
				$\xymatrix@C=5ex@R=4ex{		
					&&&
					\texttt{$\texttt{S} \ \mid \ \texttt{L''} \ \mid \ \Gamma$}
					\ar^{\{\texttt{L''} \mapsto \texttt{eps}\}}[dl]
					\ar_{\{\texttt{L''} \mapsto \texttt{1 \ L'''}\}}[dr]			
					\\	
					&&		
					\texttt{$\texttt{eps} \ \mid \ \texttt{eps} \ \mid \ \Gamma$}
					&&
					\texttt{$\texttt{S} \ \mid \ \texttt{L'''} \ \mid \ \Gamma$}
				}$
			}
			\caption{Folding variant narrowing tree for the goal \texttt{$\texttt{S} \ \mid \ \texttt{L''} \ \mid \ \Gamma$}.}
			\label{fig:graph-paser2}  
			\vspace{-1ex}
		\end{figure}	
		
		\begin{example} \label{ex:parser2}
			Consider again the elementary parser defined in Example \ref{ex:parser1} and the initial configuration $\texttt{init} \ \mid \ \tt L \mid \Gamma$. 
			Following the PE algorithm, we construct the two
 folding variant narrowing trees that 
 are shown in 
 Figures \ref{fig:graph-paser} and  \ref{fig:graph-paser2}.
Now all leaves in the tree are closed w.r.t. $Q$,  and by applying the post-partial evaluation transformation with the independent renaming $\rho = \{\texttt{$\texttt{init} \mid \texttt{L} \mid \Gamma$} \mapsto \texttt{finit(L)}, \texttt{$ \texttt{S} \mid \texttt{L} \mid \Gamma$} \mapsto \texttt{fS(L)}, \texttt{$\texttt{eps} \mid \texttt{eps} \mid \Gamma$} \mapsto \texttt{feps}\}$, we get the following specialized program 

{
	\small
			\begin{verbatim}
			    eq finit(eps)   = feps .         eq finit(1) = feps .        
			    eq finit(0 L)   = finit(L) .     eq fS(eps)  = feps .
			    eq finit(1 1 L) = fS(L) .        eq fS(1 L)  = fS(L) .
			\end{verbatim}
}			
\vspace{-1ex}
			
\noindent
		that is even more efficient and readable than the specialized program shown in the Introduction. Note that we obtain \texttt{finit(1 eps) = feps} but it is simplified to \texttt{finit(1) = feps} modulo identity.
		\end{example}
		

		\section{Experiments and Conclusions}\label{sec:impl}\label{sec:experiments}
		We have implemented the transformation framework presented in this paper. 
		We do not yet have an automated tool where you can give both a Maude program and an initial call, and the tool returns the specialized program.
		However, all the independent components are already available and we have performed some experiments in a semi-automated way,
		i.e., we make calls to the different components already available without having a real interface yet:
		equational unfolding (by using folding variant narrowing already available in Maude; see \cite{maude-manual}), 
		equational closedness (we have implemented Definition~\ref{closedness} as a Maude program), 
		equational embedding (we have implemented Definition~\ref{def:embedding} as a Maude program), 
		and equational generalization and abstraction (we have implemented Definition~\ref{abstract} as a Maude program that
		invokes a Maude program defining the least general generalization of  \cite{AEEM2014}).

		Table~\ref{tbl1} contains  the experiments that we have performed using a 
		MacBook Pro 
                 with an Intel Core i7 (2.5Ghz) processor and 8GB of memory
                 and considering the average of ten executions for each test. 
		These experiments are available at  \url{http://safe-tools.dsic.upv.es/victoria}.
		We have considered the three Maude programs discussed in the paper: 
		Parser (Example~\ref{ex:parser1}), Double-flip (Example~\ref{ex:flip}), and Flip-fix (Example~\ref{ex:graph}), and
		three sizes of input data: one hundred thousand elements, one million elements, and five million elements.
				Note that elements here refer to 
		graph nodes for Double-flip and Flip-fix, and list elements for Parser.
We have benchmarked three versions of each program  on these data: original program, partially evaluated program (before post-processing renaming), and final specialization (with post-processing renaming).
The    relative speedups that are achieved thanks to specialization are given  in the {\sf Improvement} column(s) and computed as the percentage 
 $\sf 100 \times (OriginalTime - PETime)/OriginalTime$.
		For all the examples,  the partially evaluated program has a significant improvement in the execution time when compared to the original program, both with and without renaming, but more noticeable after  renaming.  Actually,   matching  modulo axioms such as associativity, commutativity, and identity are pretty expensive operations that are massively used in Maude,
		and  can be 
		drastically reduced after specialization (i.e., the Parser example moves from a program with ACU and Ur operators to a program without axioms). 

\vspace{-0.05cm}

Developing a complete partial evaluator for 
the entire Maude language requires to deal with some   features not considered in this work,
and to experiment with refined 
heuristics that maximize the specialization power.
		Future implementation work will focus on automating
		the entire PE process for a large subset of the language including conditional rules, memberships, and conditional equations. 
		This, in turn,
will necessitate some new developments
in the Maude narrowing infrastructure.
In this sense,  advancing 
the present PE research ideas
will be a significant driver of 
new symbolic reasoning features in Maude.

\begin{table}[t]
	\centering
	\begin{tabular}{lc|r|r|r|r|r|}
		\cline{3-7}
		&& \multicolumn{1}{c|}{\textbf{Original}} & \multicolumn{2}{c|}{\textbf{PE before renaming}}  & \multicolumn{2}{c|}{\textbf{PE after renaming}}                                       \\ 
		\hline
		\multicolumn{1}{|c|}{\textbf{Benchmark}} & \multicolumn{1}{c|}{\textbf{Data}} & \multicolumn{1}{c|}{\textbf{Time (ms)}}    & \multicolumn{1}{c|}{\textbf{Time (ms)}} & \multicolumn{1}{c|}{\textbf{Improvement}} & \multicolumn{1}{c|}{\textbf{Time (ms)}} & \multicolumn{1}{c|}{\textbf{Improvement}} \\ \hline
		\multicolumn{1}{|l|}{Parser}    & 100k         & 156           & 40              & 74,36          & 35         & 77,56                                     \\ \hline
		\multicolumn{1}{|l|}{Parser}    & 1M         & 12.599       & 418             & 96,68            & 361       & 97,13       \\ \hline
		\multicolumn{1}{|l|}{Parser}     & 5M        & 299.983   & 2.131      & 99,29         & 1.851                & 99,38              \\ \hline
		\multicolumn{1}{|l|}{Double-flip} & 100k       & 177            & 155            & 12,43          & 86         & 51,41                                     \\ \hline
		\multicolumn{1}{|l|}{Double-flip}  &1M      & 1.790        & 1.584           & 11,51           & 871         & 51,34          \\ \hline
		\multicolumn{1}{|l|}{Double-flip}   &5M     & 8.990     & 8.006      & 10,95        & 4.346                  & 51,66               \\ \hline
		\multicolumn{1}{|l|}{Flip-fix}    & 100k       & 212      & 188           & 11,32           & 151        		& 28,77                                      \\ \hline
		\multicolumn{1}{|l|}{Flip-fix}    &1M       & 2.082     & 1.888         & 9,32            & 1.511         	& 27,43             \\ \hline
		\multicolumn{1}{|l|}{Flip-fix}    &5M       & 10.524    & 9.440         & 10,30          & 7.620            & 27,59              \\ \hline
		\newline
	\end{tabular}
\vspace{-2ex}
	\caption{Experiments}
	\label{tbl1}
\vspace{-6ex}
\end{table}

\end{document}